\documentclass[10pt]{article}
\pdfoutput=1

\usepackage[english]{babel}
\usepackage[utf8]{inputenc}
\usepackage{amsmath,amssymb}
\usepackage{graphicx}
\usepackage{subfigure}

\title{The chaotic set and the cross section 
for chaotic scattering in 3 degrees of freedom}
\author{C Jung$^1$, O Merlo$^2$ T H Seligman$^{1,3}$ and W P K Zapfe$^1$}

\begin{document}

\maketitle

$^1$ Instituto de Ciencias F\'isicas \\
Universidad Nacional Aut\'onoma de M\'exico, \\
Av. Universidad s/n, Apdo. Postal 48-3
Cuernavaca, Morelos, M\'exico.\\
$^2$ Zurich University of Applied Sciences \\
Institute of Applied Simulation \\
Grüental, P.O. Box \\
CH-8820 Waedenswil, Schweiz. \\
$^3$ Centro Internacional de Ciencias AC \\
Apartado Postal 6-101 \\
C.P. 62131 Cuernavaca, Morelos, M\'exico.

\begin{verbatim}
karelz@fis.unam.mx
\end{verbatim}

\bibliographystyle{unsrt}

\begin{abstract}

    This article treats chaotic scattering with three degrees 
of freedom, where one of them is open and the other two are 
closed, as a first step toward a more general understanding of
chaotic scattering in higher dimensions.
 Despite of the strong restrictions it breaks 
the essential simplicity implicit in any two-dimensional 
time-independent scattering problem. 
Introducing the third degree of freedom by breaking a 
continuous symmetry,
we first explore the topological structure of the 
homoclinic/heteroclinic tangle and the structures in the scattering 
functions. Then we work out implications
of these structures for the doubly differential cross section.
The most prominent structures in the cross section are rainbow 
singularities. They form a fractal pattern which reflects the
 fractal structure of the chaotic invariant set. This allows to 
determine structures in the cross section from the invariant 
set and conversely, to obtain information about the topology of 
the invariant set from the cross section. 
The latter is a contribution to the inverse scattering problem 
for chaotic systems.
\end{abstract}

PACS Numbers 05 45

\newpage

 \section{Introduction}

Chaotic scattering with two degrees of freedom is fairly well understood 
\cite{cha, jus, js1, eck, rapoport, mitchdelos} 
both for hard chaos (hyperbolicity) and soft chaos 
in terms of Smale's horseshoe construct \cite{sma, butik}. 
In a time-independent Hamiltonian system with two 
degrees of freedom, 
Smale's horseshoe displays the invariant manifolds 
which qualitatively separate the dynamics. On an appropriate 
surface of section these invariant sets become
smooth curves. This has two advantages:1) we can study
objects easily if they can be drawn on paper, and 2)
they separate the phase space dynamics if they 
close or go to infinity in both directions. 

Scattering functions \cite {butik,tapia, junselmer} 
and cross sections 
\cite{pue, schelin} have been analysed both statistically and geometrically,
and the inverse scattering problem has been tackled \cite{tapia,jls}
(for general background information see
the text books \cite{ram,gla,aks}).

Now the effort must be directed towards higher dimensional systems,
as they are relevant in astrophysics \cite{contop, wiggescape}
or chemistry \cite{waalkensgeo, ezra, waalkensstate}.
How do we proceed to generalise this success in the description 
of chaotic scattering to more than two degrees of freedom?
There are previous important steps particularly by Wiggins 
\cite{wi2,wig, wigginshill}, 
but also by Ott \cite{chen}. The former develops a formal theory that is 
difficult to apply but gives important foundations of the problem.
The latter show, that a straight forward generalisation of the 
three-disc system in triangular configuration to a four sphere 
problem in tetrahedral configuration is feasible, 
but the invariant manifolds are low dimensional and thus not
relevant so some physical problems, as they are difficult to detect.  
We have suggested in an 
earlier paper that it can be useful to have a gradual approach 
to a change in dimension by breaking a continuous symmetry of the
problem \cite{bz}. 

In the present paper we shall follow  this last 
suggestion and apply it to the simplest possible three degrees of
freedom 
configuration, with two bound and only one open degree of freedom. 
Despite of this strong restriction, we are now definitely beyond
 the situation, where Smale's theory can be used,
 and systems of that kind are not without interest. 
We might consider guiding channels with 
one open degree of freedom, or we can consider 
equivalently three interacting particles in one dimension, 
under conditions where asymptotically one pair must always 
be bound.

 We shall focus on the topological structure of the invariant 
set rather than on details of trajectories, because the former 
properties tend to be generic , i.e. robust  and 
embody the most relevant information about the chaotic 
system \cite{jls}. We are thus looking for some generalisation 
of the horseshoe to higher dimensions. A simple minded
generalisation does not exist, as can be seen from the work 
of Wiggins \cite{wig}. The basic reason is that the dimension
of the invariants manifolds of hyperbolic points is too low.

    We proceed by analysing the transition from a system, 
which has an ${\cal O}(2) $ symmetry group 
in the three dimensional position space, 
reducing the problem for a fixed value of the angular momentum
 to a system with two degrees of freedom. 
We shall then introduce in the reaction region 
(i.e. in the region where the asymptotic Hamiltonian is not valid)
 a symmetry breaking term. This will allow us to start our 
analysis with a continuous stack of horseshoes corresponding to 
different values of the angular momentum
 and observe how the structure evolves as the 
invariance breaks down. We then study the implications for 
scattering functions and for the doubly differential cross 
sections. The fractal structure of singularities in the latter,  
i. e. the rainbow structures, is directly related 
to the structure of the chaotic set, leading to an additional important  
contribution to the corresponding inverse scattering problem. 
Indeed we shall also see, that the basic idea developed is not 
limited to three degrees of freedom nor to one open degree of 
freedom, but these aspects will only be commented.

    We shall illustrate our procedures with three examples. 
First a channel with harmonic walls in two directions that 
deforms to some more complicated potential in some compact 
region of configuration space. This is equivalent to two pairs
of particles bound by harmonic forces and
interacting between each other weakly. 
The second is a bottle shaped billiard that is either 
connected to a channel billiard or has an opening that 
separates the interior from the exterior in some plane in 
configuration space. Finally for numerical convenience we 
shall use a $\delta $ kicked system of two degrees of freedom, 
which is topologically 
equivalent to the Poincar\'e maps of the first system.

    As far as the physical implications are concerned, it 
does not seem easy to emulate systems with such restrictions 
and no significant friction, though it might become feasible 
with Bose Einstein condensates at some point in the future.
The alternatives in celestial mechanics imply higher 
dimensions and/or more complicated asymptotes, as is the case 
with the breaking of the Jacobi invariant in ref \cite{bz}. Yet 
the results are certainly useful for semi-classical 
considerations, such as e.g. the ones used in \cite{papen} to 
design and interpret a microwave experiment.

    The paper is structured as follows: in the next section 
we shall define the class of three degree of freedom systems, 
which we shall consider as well as the specific examples we
 shall use. Next we proceed to construct the invariant sets and 
study them as a function of the symmetry breaking. 
In section four we derive the scattering functions, 
that is the behaviour of outgoing asymptotes as a function of 
the incoming ones. This will allow us to discuss the rainbow 
effects in the doubly differential cross sections in section 
five. We finally proceed to comment on the scope of our 
results as well as implications for semi-classics.

\section{The class of systems considered and the model systems}

We consider a scattering system with three 
degrees of freedom where one of
them is open and the other two are closed. We imagine 
that the first
degree of freedom is a translational degree of freedom between projectile
and target, the second degree of freedom is a vibrational degree of freedom
and the third one is a rotational degree of freedom.

The first particular model system of this class which we use is
a rotationally symmetric channel containing an additional short range 
potential representing 
some obstacle (the target) sitting in the otherwise empty channel. 
Accordingly the total Hamiltonian of the
system splits, as always in scattering systems, in a free Hamiltonian $H_0$
describing the asymptotic free motion ( in this particular case it is the 
motion in the empty channel), and an additional interaction
$W$ which is the short range scattering potential, i.e.

\begin{equation}\label{hamgen1}
H = H_0 + W
\end{equation}

We imagine that the channel runs straight in one direction and we use the
coordinate $q$ along the channel, $p$ is the momentum conjugate to $q$.
In the transverse direction the motion of the projectile is confined by the
channel and we assume that the potential which represents the walls of the
channel is quadratic in the transverse coordinates. Because of the rotational
symmetry of the empty channel one natural choice is 
to use cylindrical coordinates $\rho$ and $\theta$ and 
their conjugate momenta $p_{\rho}$ and $L$
and also to include into the free Hamiltonian the quadratic potential and the 
transverse kinetic energy in these coordinates. 
Since the radial motion is
an oscillation, it is sometimes more convenient to use action and
angle variables $I$ and $\phi$ for the radial degree of freedom.
We will switch freely between these two possibilities according to which one
is more convenient at the moment.  
The asymptotic Hamiltonian written in cylindrical position 
and momentum coordinates
results as
\begin{equation}\label{hamgen2}
H_0 = p^2/2 + p_{\rho}^2/2 + L^2/(2 \rho^2) + \rho^2/2
\end{equation}
We choose units of time such that the transverse oscillation frequency in the
empty channel is one. To describe the obstacle we use later as model 
of demonstration the potential 
\begin{equation}\label{hampert}
W = -\frac{\exp(-D)}{D}
\end{equation}
where
\begin{equation}
D^2 = q^2+\rho^2(\sin^2 \theta +(1+A)^2\cos^2\theta)+1
\end{equation}
The extra constant, taken as one, avoids singularities in the
potential.
Very important for the 
following is the parameter $A$ which measures the deviation of the
interaction potential from rotational symmetry. For $A=0$ we have perfect 
symmetry and since the free Hamiltonian $H_0$ is also symmetric, the total
system is symmetric, the angular momentum $L$ is conserved and the system
can be reduced to two degrees of freedom. For the value $L=0$ the symmetric 
system reduces to the 
one used in \cite{pue}. 
This reference also contains a physical interpretation
of similar potential models.

Next we need labels for the asymptotes of the system, i.e. the 
trajectories of the free motion described by $H_0$. The optimal possibility 
is to use 5 independent quantities which all are constant along the 
asymptotic trajectories.
Since in any Hamiltonian system the energy 
$E$ is conserved, both under the free asymptotic dynamics and under the full
dynamics, we use $E$ as one of these labels. 
Asymptotically the translational degree of freedom decouples from the other
ones and the momentum $p$ becomes constant. We use $p$ as second asymptotic
variable. The sign of $p$ indicates into which direction the asymptotic 
motion runs and therefore a comparison between the signs of initial and final
$p$ distinguishes transmission and reflection. 
In addition, since $H_0$ is
independent of $\theta$, the conjugate variable $L$ is conserved under the
asymptotic dynamics and it is convenient to use it as third label for
asymptotic trajectories. To distinguish the various asymptotic trajectories
with the same values of $E$, $p$ and $L$ we need two additional labels 
giving the relative
phase shifts between the translational motion along the channel and the
other two motions. 
The systematic choice for these relative angles are the reduced phases
constructed along the following idea. In action variables we find for our
particular system
\begin{equation}
H_0 (p,I,L)= p^2/2 + 2I + |L|
\end{equation}
Then the asymptotic equations of motion for $\phi$ and $\theta$ are
\begin{equation}\label{phidt}
\frac{d \phi}{dt}= \frac{\partial H_0}{\partial I} =  \omega_\phi(I,L) = 2
\end{equation}\
\begin{equation} \label{thpunto}
\frac{d \theta}{dt} = \frac{\partial H_0}{\partial L}= \omega_\theta(I,L) = \pm 1
\end{equation}
The sign appearing in equation \ref{thpunto} is the sign of the angular momentum.
We define the reduced angles $\psi$ and $\chi$ belonging to $\theta$ and
$\phi$ respectively as
\begin{equation}\label{psilabel}
\psi = \theta - \omega_\theta(I,L) q/p = \theta \mp q/p
\end{equation}
\begin{equation}
\chi = \phi - \omega_\phi(I,L) q/p = \phi - 2q/p
\end{equation}
The sign in equation \ref{psilabel} is minus the sign of the angular momentum.
A short calculation shows that these two reduced angles are
constant under the asymptotic dynamics.

An equivalent possibility is the following: We use the values of the 
two coordinates $\phi$ and $\theta$ at the moment when the absolute value
of $q$ reaches some very large value (in the numerical examples, we use 
 $\|q\| = 8$). We call these two particular values which serve as asymptotic 
labels $\chi$ and $\psi$ again.
We use the index $in$ for initial asymptotic conditions and
the index $out$ for the labels of outgoing asymptotes.
 
Since it is simpler to handle iterated maps
instead of flows, we will use Poincar\'e maps to represent the dynamics
of the system and to explain many ideas (an elementary explanation of the 
concept of Poincar\'e maps is found in section 2.5 of \cite{poi}). 
For the channel problem an
appropriate intersection condition for the Poincar\'e map is to take the
maximum of the cylindrical radial coordinate $\rho$.
This choice has the advantage to coincide with the 
choice made for a billiard model 
which we also use later.
Almost all trajectories intersect this surface
transversely an infinite number of times and asymptotically the return time
to this surface becomes constant, 
note the constant value of $\omega_\phi(I,L) = 2$
in equation \ref{phidt}. 
The only exceptions are
trajectories with action $I=0$ which run along spirals of 
constant value of $\rho$
without any radial oscillations. In the domain of the map we use the
canonical coordinates $q$, $p$, $\theta$, $L$.

An alternative version of the reduced angles $\psi$ and $\chi$ 
for the map is obtained as follows: For trajectories with the value p of the
asymptotic longitudinal momentum select an interval of large 
absolute values of $|q|$, 
say the interval $[Q,Q+|p|]$ where $Q$ is sufficiently 
large to guarantee that the
trajectory is already in the asymptotic region. Observe 
the trajectory of the map 
until it steps into this interval. For this point define 
$\chi=2 \pi (|q|-Q)/|p| $ 
and for $\psi$ take the actual value of $\theta$ in this point. 
Note that the initial point and the final point of the selected 
$q$ interval can
be identified for the purpose of asymptotic labelling 
since they describe the
same trajectory (one is the image of the other point under 
the action of the map)
and lead to the same value of the angle $\chi$ since multiples of $2 \pi$ are
irrelevant. Therefore this line of initial 
conditions has the topology of a circle.
An analogous construction also works well for the Poincar\'e 
maps of other model systems.

The second model of demonstration used is a billiard model, it describes 
scattering in a bottle. 
The boundary of the bottle will be defined with the 
aid of the following functions:

\begin{equation}\label{border}
f(q)=\begin{cases}
1.4\sqrt{1-q^2}, &\hbox{ for } q\leq 0\\
a_0 q^{2.5}+a_1 q^2+1.4, &\hbox{ otherwise.} 
\end{cases}
\end{equation}

\begin{equation}
r(q,\theta)=1+ A \cos(\theta) \cos (\frac{q\pi}{2 q_0})^2
\end{equation}

The boundary itself is given by 
\begin{equation}
\rho(q,\theta)=r(q,\theta)f(q)
\end{equation}

Two dimensional billiards with the border described by function 
\ref{border}  have been studied in \cite{echo} and 
a similar model in \cite{hansen}.

The values of the constants $a_i$ are chosen to give a
smooth boundary and to lead to an unstable quasi periodic orbit 
in the plane $q=0$ so that there is a complete binary horseshoe for
the reduced rotationally symmetric case for $L=0$. 
Convenient values turn out to be:

\begin{align}
a_0 &=\frac{14}{25}(3.49)^{-1/4}\\
a_1 &= -0.7\\ 
q_0 &= (3.49)^{1/2}
\end{align}
 
Also this model has a parameter $A$ which gives the distance from rotational
symmetry and again for $A=0$ we have perfect symmetry and the system can be
reduced to one with two degrees of freedom.

The billiard dynamics is the usual one with elastic reflection
on the wall and free motion in the inside. 
The energy is a constant and proportional to the square of the
momentum. Therefore we can set the velocity $\|v\|=1$ without 
loss of generality.
 The Poincar\'e map of the bottle
model is the usual Birkhoff map on the wall of the billiard 
(these are explained
in \cite{markchern}).
We only use weak deformations of the bottle such that this map coincides
with the intersection condition of maximal cylinder radius $\rho$.
Accordingly we use in the domain of the map the same coordinates
$q$, $p$, $\theta$ and $L$ as in the channel model.

 The Birkhoff-Poincar\'e map of this system has a binary horseshoe instead
of the ternary one of the other two examples, and it has a 
surface of no return 
at a finite value of $q$, which makes is qualitatively different. 
Nevertheless we will see that it belongs to the same class of systems.  

Since it is a lot more convenient and faster to investigate maps instead 
of flows we construct as third model a closed form example of a map acting
on the 4 dimensional domain with coordinates $q$, $p$, $\theta$ and $L$.
This model is constructed in a way that it can serve as a prototype
Poincar\'e map for the whole class of systems considered in this article.
 It is based on the usual scheme of kick and free flight. 
The generating function for the kick is:

\begin{equation}\label{genmap}
G(q,\theta, \tilde{p},\tilde{L})=
q\tilde{p}+\theta\tilde{L}+(L_{max}-\tilde{L})(1+A cos\theta)V(q),
\end{equation}
with potential function

\begin{equation}\label{potmap}
V(q)=-e^{-q^2}.
\end{equation}

We define the force function on the $q$ coordinate as
\begin{equation}\label{fenq}
F(q)=-dV/dq=-2qe^{-q^2}.
\end{equation}

And the kick map is implicitly given by:

\begin{align}
\tilde{q} &= \frac{\partial G}{\partial \tilde{p}} &
p &= \frac{\partial G}{\partial q} \\
\tilde{\theta} &= \frac{\partial G}{\partial \tilde{L}} &
L &= \frac{\partial G}{\partial \theta}.
\end{align}

To construct a complete step of the map, the particles will perform first 
 half of the free flight:

\begin{align}
q' &=q_n+ p_n/2\\
\theta'&= \theta_n + L_n/2\\ 
p' &=p_n\\
L' &=L_n 
\end{align}

Afterwards, we give this auxiliary coordinates the kick, given explicitly 
by the next set of equations:

\begin{align}
q'' &=q'\\
\theta''&= \theta'-(1+A\cos \theta')V(q') \\
p'' &= p'+(L_{max}-L')\frac{(1+A \cos \theta') F(q')}
{1+AV(q')\sin\theta'} \\
L'' & =\frac{L'+L_{max}A V(q')\sin\theta'}{AV(q)\sin\theta'+1}
\end{align}

And finally half a free flight is applied again:

\begin{align}
q_{n+1} &=q''+ p''/2\\
\theta_{n+1}&= \theta'' +  L''/2\\ 
p_{n+1} &=p''\\
L_{n+1} &=L'' 
\end{align}

This concludes the action of the map. Every step is symplectic,
and the map behaves qualitatively as the Poincar\'e map of the 
first example. 

We use mainly the map model for the presentation and explanation 
of our ideas.
Below we compare some results of this map with the results of the other 
two models in order to convince ourselves that the map
 results are representative.

\section{The topological structure of the chaotic invariant set}

Since the interaction potential of equation \ref{hampert} or of equation \ref{potmap}
is negative, there are trajectories
with negative energy yet arbitrarily close to zero, going out extremely far 
and returning.
Therefore our first example system has no points without return. 
Its outermost 
periodic orbits lie at $q = \pm \infty$. They are trajectories oscillating
transverse to the channel and rotating at constant value of $q$,
where $q$ is
arbitrarily far away. Since any displacement of $q$ leads to an equivalent 
transverse oscillation, these transverse trajectories come in a whole
continuum of copies. In the Poincar\'e map they lead to fixed points at
$q =\pm \infty$, $p=0$ and these points are parabolic, 
i.e. neutrally stable in linear
approximation. However the non-linearities of the map at finite
values of $q$ make these points non linearly unstable 
and they have stable and unstable
invariant manifolds. The stable manifolds consist of trajectories which go 
out to infinity
monotonically, i.e. the absolute value of $q$ increases monotonically while
at the same time the value of the momentum $p$ converges to zero.
Asymptotically all energy of the motion goes to the transverse motion.
The unstable manifolds consist of trajectories doing the same under the
time reversed motion. In this sense the trajectories belonging to the 
invariant manifolds of the points at infinity converge to the periodic
trajectories described above. In the domain of the map
these manifolds are the separatrix lines between motion going  
monotonically away and motion which returns. 

The map model described as third example of demonstration in the previous
section is in this respect similar to the channel. It also does not
have values of $q$ of no return.

In contrast the bottle model has a surface of no return, the surface of 
the bottle neck. The trajectories lying on this surface forever serve as
outermost fixed points in the Poincar\'e map.

The set of trajectories staying in the plane of the bottle neck and also the
trajectories with $p=0$ in the asymptotic region of the channel form an
example of what Wiggins \cite{wig} calls a normally hyperbolic invariant
manifold, abbreviated NHIM. Let us explain this set first 
for the case of the
bottle, since in this case we have a NHIM in its original form. 

For fixed energy, all the orbits which stay forever 
in the plane of the bottle neck form a
two dimensional continuum. First, there are such orbits for all possible
 values
of the angular momentum and second to each such orbits there also exist
corresponding orbits rotated by an arbitrary angle. 
These trajectories are neutrally stable
under perturbations of initial conditions which keep them in the bottle
neck plane. In the domain
of the Poincar\'e map all these trajectories form a 2 dimensional surface,
which is the NHIM in the domain of the map.
In the direction perpendicular to the bottle neck all these trajectories are
unstable and they have stable and unstable manifolds. The union of these
invariant manifolds taken over the whole 2 dimensional continuum of bottle
neck trajectories form the stable and unstable manifolds of the NHIM
which we will call $W^s$ and $W^u$.
They are 3 dimensional surfaces in the 4 dimensional domain of the map.
Therefore these surfaces are dividing surfaces, they divide trajectories
which pass the bottle neck from trajectories which return, for more details 
and consequences for scattering see \cite{wi2}, or for  applications
see \cite{wigginshill}. 

In the example of the channel 
we have a slightly modified version of a NHIM. 
Trajectories in the empty channel or in the asymptotic
 region of the channel 
having $p=0$ stay 
forever in the plane of a fixed value of $q$, they are trajectories in the 
two dimensional oscillator potential forming the empty channel. 
In a harmonic 
oscillator all trajectories are periodic, they are ellipses. The ones for a 
fixed value of the total energy are a two dimensional continuum where each 
individual trajectory can be distinguished by its value of the 
angular momentum
and by the orientation of the ellipse. Again such trajectories are neutrally
stable under perturbations which keep the value $p=0$. 
The difference between
the channel and the bottle example is that in the channel case the 
trajectories which form the NHIM are also neutrally stable in $q$ direction,
at least in linear approximation. If we include the 
non linearities caused by
the asymptotic tail of the attractive scattering potential, they become 
unstable and
have stable and unstable manifolds. The union of these invariant manifolds
forms again the invariant manifolds of the whole NHIM. They are
dividing surfaces which separate trajectories going out monotonically from
returning trajectories. In this respect the map model behaves like the 
Poincar\'e map of the channel model.

To investigate the chaotic set of the models we start with the case of
$A=0$ and later on break the symmetry. For $A=0$ the $\theta$ dependence
of the dynamics decouples from the rest of the dynamics, $L$ becomes a
conserved quantity and the map reduces to a 2 dimensional map in the 
coordinates $q$ and $p$, where $L$ serves as a parameter. Because of its
importance in what follows we show this reduced map in explicit form
for the map model with the potential function from equation \ref{potmap}:
\begin{align}
q_{n+1} &= q_n+p_n-(q_n+p_n/2)\exp(-(q_n+p_n/2)^2)(L_{max}-L)\\
p_{n+1} &= p_n-2(q_n+p_n/2)\exp(-(q_n+p_n/2)^2)(L_{max}-L)\\
\end{align}

\begin{figure}
  \centering
  \subfigure[$L=0.00$]{\label{lzero}\includegraphics[width=0.66\textwidth]{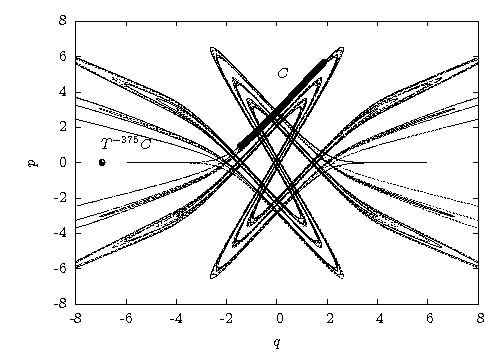}}
  \subfigure[$L=2.60$]{\label{lmed}\includegraphics[width=0.66\textwidth]{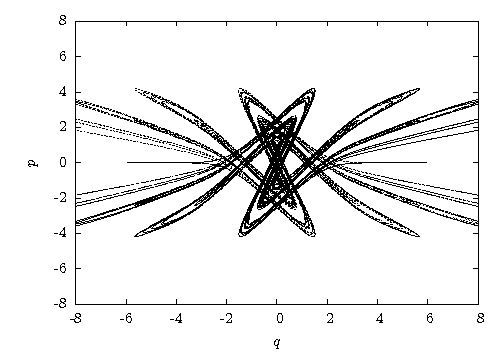}}
  \subfigure[$L=5.71$]{\label{lbig}\includegraphics[width=0.66\textwidth]{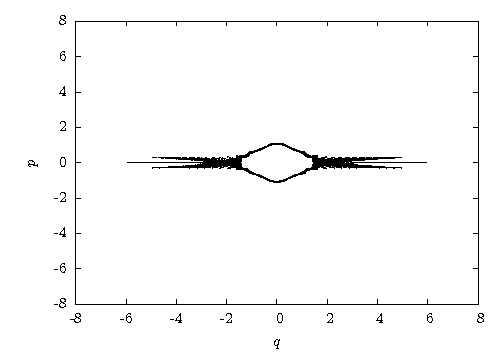}}
  \caption{The Poincar\'e Maps for the discrete dynamical system in the axial symmetric case ($A=0$). The $L$ parameter regulates the degree of development of the horseshoe. Notice also reduction of the phase
space occupied by the horseshoe as $L$ increases. The line $C$ on \ref{lzero} it's choosen so that it's preimage is an adecuate domain for scattering functions.}
  \label{poincares01}
\end{figure}

The $L$ value determines the strength of the force of the kick, becoming
negligible for $L \approx L_{max}$. This corresponds to all energy 
spent on rotational motion, and almost no development for the 
horseshoe, as can be seen on the sequence of plots of the homoclinic tangle 
for this reduced map in figs. \ref{poincares01}.

\begin{figure}
  \centering
  \subfigure[$L=0.00$]{\label{lzeroc}\includegraphics[width=0.66\textwidth]{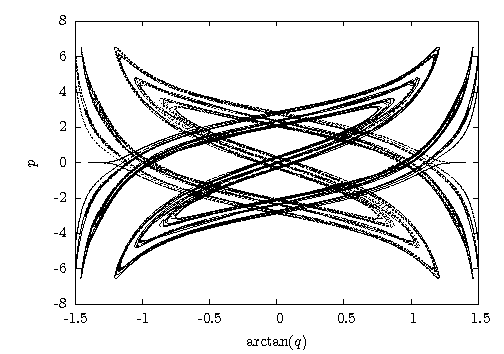}} 
  \subfigure[$L=2.60$]{\label{lmedc}\includegraphics[width=0.66\textwidth]{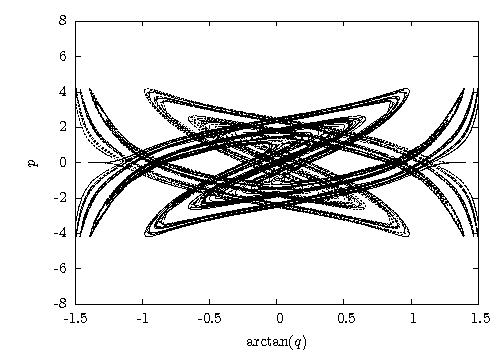}}
  \subfigure[$L=5.71$]{\label{lbigc}\includegraphics[width=0.66\textwidth]{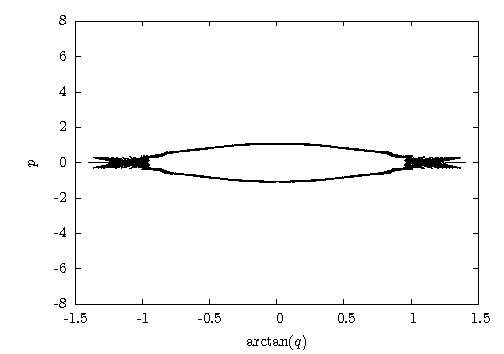}}
  \caption{Same as fig. \ref{poincares01}, 
but after compactifying the $q$ axis using the function $\arctan(q)$.  }
  \label{poincares02}
\end{figure}

To construct the horseshoe we plot the stable 
manifold and the unstable manifold of both the fixed point at $+ \infty$ 
and of the one at $- \infty$.
To plot the whole interval $q \in [-\infty, +\infty]$ on a finite range we 
use the
horizontal coordinate $z = \tanh(q)$ which is convenient to get informative
plots (figure \ref{poincares02}). The intersection points between a
 stable and an unstable manifold
are trajectories which converge forward in time and also backward in
time to a fixed point of the map. 
An intersection between the manifolds of the
same fixed point is called homoclinic point, and an intersection between the
manifolds of different fixed points is called heteroclinic point. The whole
structure of the manifolds is called 
homoclinic/heteroclinic tangle, many times only the expression homoclinic 
tangle or simply tangle is used for short even when it also contains 
heteroclinic points. The existence of a homoclinic tangle is the
topological criterion for chaos in the system. It implies the existence
of an  uncountable set of unstable trajectories. For more details on
the role of homoclinic tangles and the  importance for scattering and 
transport problems see the book by Wiggins \cite{wig}.

Because of symmetry in our example the manifolds of point $-\infty$
are obtained from the ones of the point at $+\infty$ by inversion 
about the origin. The time reversal symmetry in the maps permits also
to obtain the stable manifolds from the unstable ones by the reflection 
$p \rightarrow -p$.
From those symmetry properties also follows the existence of a trajectory
oscillating transverse to the channel at $q=0$. 
It leads to a fixed point of 
the map at $q=0$, $p=0$. We call this fixed point the inner fixed point.
In total the map has three fixed points where the outer two are symmetry
related and therefore we call the resulting homoclinic structure 
a ternary symmetric horseshoe. 

When changing $L$ we have the usual development scenario starting from a
complete horseshoe for $L=0$ up to a parabolic line when $L$ reaches
its maximal limiting value $L_{max}$ allowed by total energy. 
In our particular
example we have
\begin{equation}
L_{max} = 6.23
\end{equation}

Next we define the fundamental area $R$ for the horseshoe. Let us start
with the local branches of the manifolds of the two outer fixed points
and let them grow longer continuously in a completely symmetric pattern.
Then at some length we obtain the first intersection points, for symmetry
reasons they are located at $q=0$. 
At this moment let us stop the process
of growth. The result is a curvilinear quadrilateral which we call the
fundamental rectangle $R$.

The intersection pattern between some
local segment of an unstable manifold (e.g. the piece between an outer 
fixed point
and the next corner of $R$) with the whole fractal bundle of
stable manifolds already contains the complete information of the
whole tangle formed by the manifolds of the outer fixed points,
 and the knowledge of such local intersection patterns is
sufficient. We will make use of 
this idea for the reconstruction of important
properties of the tangle from scattering data. The use of these for
determining the development of the horseshoe is profusely explained in
\cite{jls}.

In figure\ref{poincares01} a)  we included an additional 
thick line labelled $C$ which runs parallel
to one of the local segments of unstable manifolds but just outside of $R$.
Note that the intersection pattern between $C$ and the bundle of segments of 
the stable manifolds coincides exactly with the intersection pattern between
the local segment of the unstable manifold and the bundle of segments of
stable manifolds. Accordingly also the intersection pattern along line $C$
contains all important information about the whole tangle. 
The iterated preimages of the line $C$ play an important role in the following
as asymptotic initial conditions for scattering trajectories.

Now we will discuss
the chaotic structure of the three degrees of freedom system. We begin with
$A=0$. Then the surfaces of the various values of $L$ are
dynamically independent and we get a fractal 
structure in the full dimensional
domain of the map with its coordinates $q$, $p$, $L$ and $\theta$ in a two
step process. First we pile up the tangles for the various values of $L$
obtaining a fractal tangle in a three dimensional 
embedding space and second we 
form the Cartesian product of this object with a circle representing the
fourth coordinate $\theta$.

For the general three degrees of freedom case we need an argument of
robustness of this structure under moderate deformations and moderate
breaking of the rotational symmetry. This argument follows from generic 
transversality properties.

\begin{figure}
\begin{center}
\includegraphics[width=0.7\textwidth]{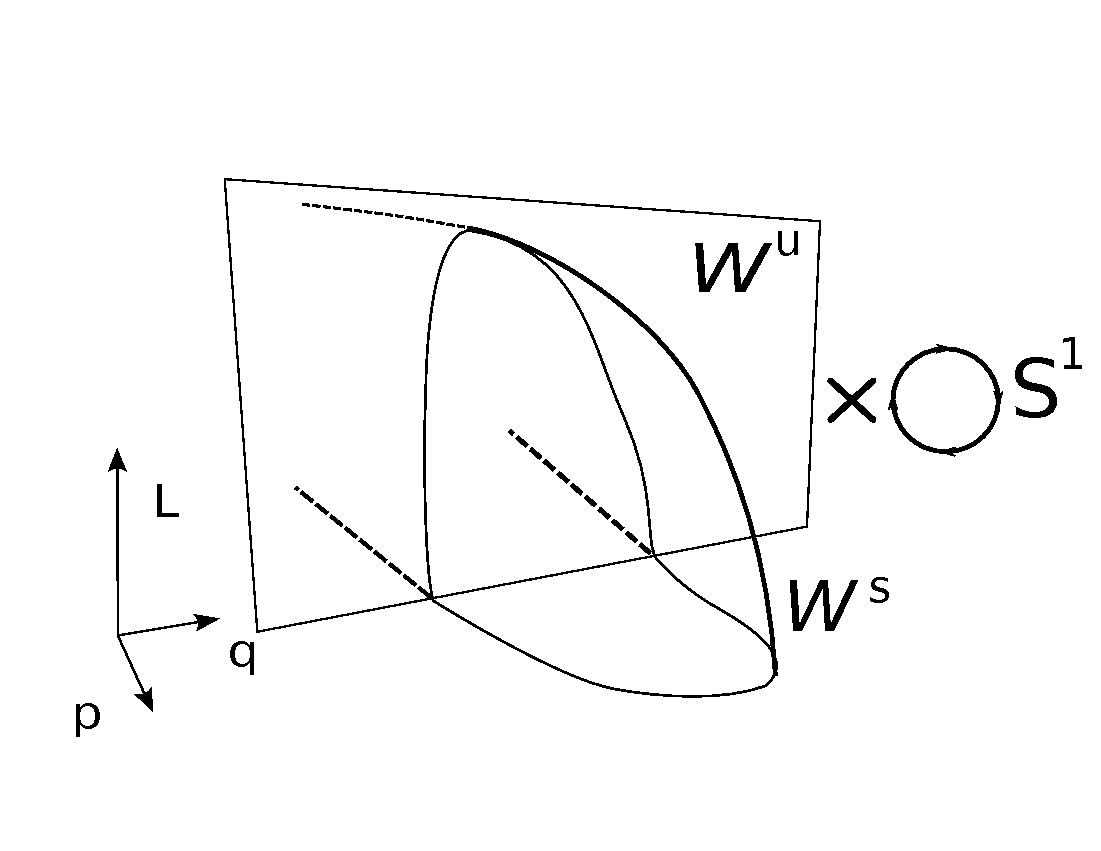}
\caption{A plot of the intersection of stable and unstable manifolds in the
$(p,q,L)$ space when $A=0$. The $L$ value gives the degree of development 
of the horseshoe for every $(q,p)$ plane. After stacking these in order,
we obtain two intersecting two dimensional manifolds. In the end we consider
a direct product with a circle for the $\theta$ coordinate.}  
\label{sabanas01}
\end{center}
\end{figure}

In a system of the form described by the equations \ref{hamgen1} and 
\ref{hamgen2}, or in the map we obtain a family 
of horseshoes as depicted in figure
\ref{poincares01} in the symmetric case. The development of 
the horseshoe is indicated by
a parameter $\gamma$ which measures the degree to which 
the horseshoe is developed.
$\gamma=0$ is the parabolic line and $\gamma=1$ is 
the complete ternary symmetric
horseshoe, for more explanations and schematic pictures see \cite{jls}.
In our present case $\gamma$ is a monotonic function of $L$.  
Accordingly also the value of $L$ 
orders the pile of two dimensional horseshoes
naturally from the complete case, with $L $ near zero, to the
integrable system, with $L$ near $L_{max}$. When we view this
family of systems as just one geometrical 
object then the family of invariant 
manifolds form smooth two dimensional surfaces in the three 
dimensional embedding space with
coordinates $q$, $p$ and $L$. For short we call this structure 
the three dimensional tangle. Now let us restrict our attention to
intersections between the unstable and stable manifolds.
The basic pattern is a transverse intersection between two two dimensional
surfaces in a three dimensional embedding space as 
shown schematically in figure
\ref{sabanas01}. 
A point of interest is that the tangential intersections of the
lower dimensional horseshoe become extremal points 
in $L$ direction of the intersection curve shown in figure \ref{sabanas01}. 
The whole curve belongs to the ``hyperbolic component'' 
of the invariant set.

So far we have ignored the $\theta$ coordinate. 
To include it for the
symmetric case of $A=0$ we simply form the Cartesian product of the
structures described above with a circle 
representing this cyclic coordinate.
Thereby the manifolds become three dimensional 
surfaces in the four dimensional
embedding space. The elementary transverse intersection 
structures between the 
surfaces $W^u$ and $W^s$ are smooth two dimensional surfaces.
Each one of their points has
one dimensional unstable and stable directions and 
two neutral directions.
Because the manifolds themselves are folded to fractal structures we obtain
a fractal repetition of the elementary intersection structures which we call
the tangle in the four dimensional domain of the Poincar\'e map. The dimension
of this tangle can be anything between two and four depending on how exactly the
manifolds turn and fold. For an argument we present later, remember that this
dimension is larger than two.

Finally we consider the changes implied by a moderate breaking of the
rotational symmetry. To understand this let us return to figure 
\ref{sabanas01} and
imagine its product with a circle. Then we see the transverse intersection
of two three dimensional surfaces in a four dimensional embedding space.
Because the intersection is transverse it is structurally stable.
This means, under small deformation the intersection pattern remains
qualitatively the same. This argument
applies to all the transverse homoclinic intersections appearing in the
development scenario of the horseshoe. However, it does not apply to the
non hyperbolic structures near the surface of KAM islands.

So, when the parameter $A$ is slightly different from
zero, we can understand that relevant qualitative 
properties of this structure 
remain unchanged. The structure of the whole four dimensional
tangle is therefore robust. This stability, together with
the transversality of the intersection, is inherited from the leaves of the 
symmetric system in a way which is only possible 
for systems which can be connected
continuously to the symmetric systems and which are not far from the 
symmetric case. This may appear restrictive at first, 
but even so we can cover a broad range of physically relevant problems.

\section{Scattering functions}

Scattering functions are interesting objects in their own right and we need 
them as auxiliary concepts to 
explain and motivate the ideas on the singularities in the cross sections,
which are more accessible to experiments.

A scattering function gives one of the outgoing asymptotic labels, 
or several 
of them, as function
of initial asymptotic conditions. It is advantageous to study outgoing 
action like 
variables as function of initial angular variables, 
this is exactly the version
of the scattering functions we need to understand the properties of the
cross sections.
As  explained before, 
every asymptotic trajectory in the Poincar\'e map
can be labelled by $p$, $L$, $\psi$ and $\chi$. It is always understood that
the total energy $E$ is kept fixed at one particular value.

The scattering function which we will now describe in detail gives
 the outgoing
momentum $p_{out}$ and the angular momentum transfer 
$\Delta L = L_{out} - L_{in}$
as function of $\chi_{in}$ and $\psi_{in}$ 
for fixed values of $p_{in}$, $L_{in}$
and $E$. The domain of this function is the 2 dimensional torus with
coordinates $\chi_{in}$ and $\psi_{in}$. Periodicity of the function
 follows thereby.

First let us study the symmetric case $A=0$ and show some numerical examples
for the model map.
In this case the angle $\psi$ is irrelevant and because 
of angular momentum
conservation the function $\Delta L$ is identically zero. 
Then the domain
of the scattering function is an interval of length $2 \pi$ 
in the angle $\chi_{in}$
and this interval is represented by the line $C$ in figure \ref{lzero}. 
Remember that $C$ is
the iterated image of an asymptotic line of fixed $p$ where $q$ 
changes over an
interval of length $p$ which corresponds to an interval of $\chi_{in}$ 
of length $2 \pi$.
  
As a consequence the scattering function has singularities on 
a fractal set and 
intervals of continuity in between. The singularities correspond the 
intersection of the line of initial conditions with stable manifolds. Since 
preimages of intersections with invariant manifolds are again 
intersections with 
the same manifolds, the intersection pattern of this line of initial 
conditions
coincides with the fractal pattern of intersections 
of line $C$ with the stable
manifolds of the horseshoe. If the asymptotic part of the trajectory starts 
on a stable 
manifold of the horseshoe and the actual scattering trajectory converges
to a periodic orbit and does not come out of the interaction region with
a longitudinal kinetic energy larger than zero, then the outgoing asymptotic
conditions are ill defined, and therefore the scattering 
function has a singularity.
The pattern of singularities is the pattern of 
intersections of the line $C$ 
with stable manifolds and is a smooth image of the fractal pattern 
in the tangle.
In this way asymptotically obtained scattering functions 
carry the information 
on the topological structure 
of the tangle sitting in the interaction region \cite{jls}.

\begin{figure}
  \centering
  \subfigure[]{\label{plzero}\includegraphics[width=0.66\textwidth]{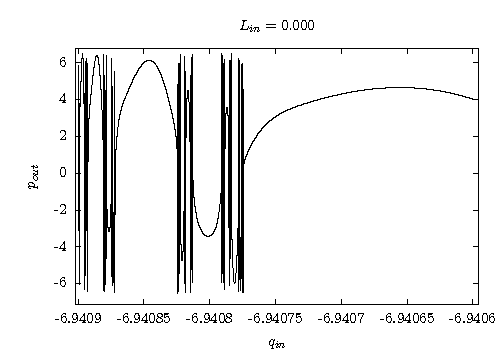}}                
  \subfigure[]{\label{plmed}\includegraphics[width=0.66\textwidth]{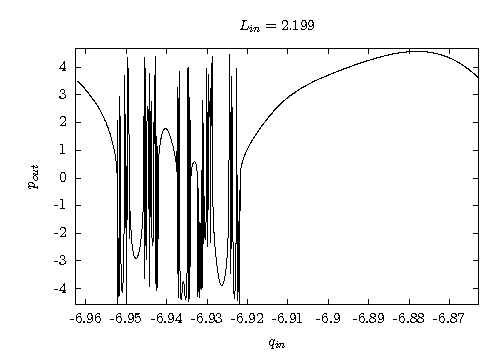}}
  \subfigure[]{\label{plbig}\includegraphics[width=0.66\textwidth]{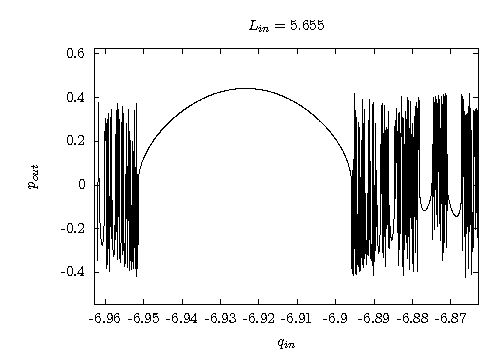}}
  \caption{The $p_{out}$ Scattering function for $A=0$ (axial symmetric case). The independent variable is $q_{in}$, the entrance phase.}
  \label{poutasym}
\end{figure}

\begin{figure}
  \centering
  \subfigure[]{\label{plow01}\includegraphics[width=0.66\textwidth]{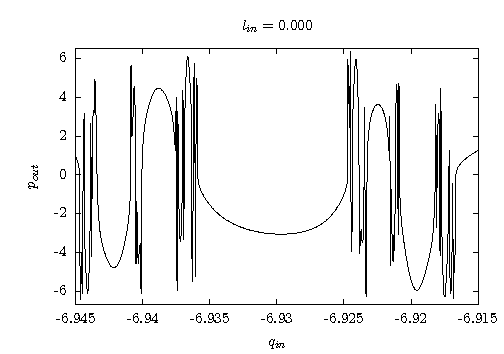}}                
  \subfigure[]{\label{plow02}\includegraphics[width=0.66\textwidth]{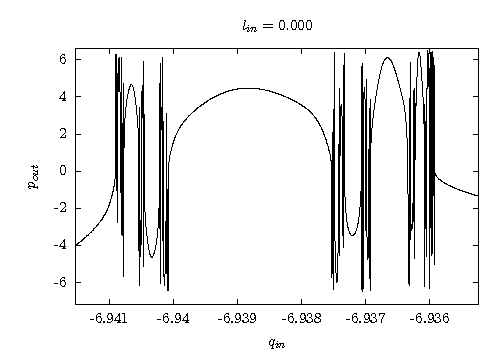}}
  \subfigure[]{\label{plow3}\includegraphics[width=0.66\textwidth]{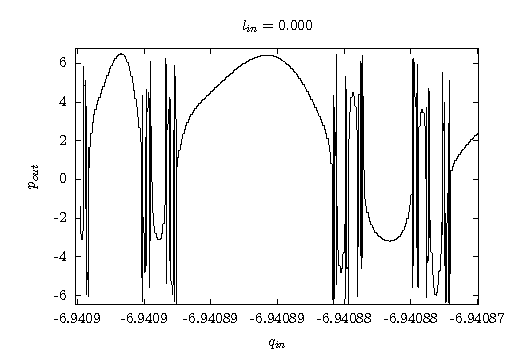}}
  \caption{An illustration of the self similarity of the scattering function $p_{out}$ with parameter $L_{in}=0.00$. This case has a complete horseshoe, 
so the function reveals the structure of  a Cantor Set in the singular values.}
  \label{poutasymzoom}
\end{figure}

We have to ensure that the line of initial 
conditions maps to a line equivalent to line $C$ in the horseshoe plots and
not to a line intersecting the outer tendril further out, where it does not
scan the whole structure. As we move line $C$  
further away from $R$ we loose more and more intersection points in 
the line, until it is so far out that it does no longer intersect the
bundle of stable manifolds at all. Here it is essential to use a value of
$p_{in}$ sufficiently small. 
The actual line used for the numerical examples of 
the scattering functions and
cross section data will be the 312th preimage of the line 
$C$ shown in the figure, 
which is sufficiently far out to be considered asymptotic. 
This line is given by
\begin{equation}\label{cpreimage} 
T^{-312} C = \{(q,p)\| q \in [-6.962,-6.912], p_{in} = 0.05 \},
\end{equation}
where $T$ is the Poincar\'e Map.
If the line of initial conditions intersects the fractal structure only
partially, then it still would contain the complete 
information because of the
self-similarity of the fractal structure. 
However, then the reconstruction would
pose some additional technical problems which we want to avoid. 
Typical structures for the scattering functions on a one dimensional domain
are presented in figures  \ref{poutasym} and  \ref{poutasymzoom}. 

In a given plot of a scattering function we recognise intersections with the 
stable manifolds as initial conditions
which lead to $p_{out}=0$. They represent boundaries between transmission
($p_{out}>0$) and reflection ($p_{out}<0$).

In preparation for the asymmetric case we have to include the role of
the coordinate $\psi$.
For $A=0$ the fractal of singularities in the two dimensional domain 
of the scattering 
function is the Cartesian product of a one dimensional fractal along the
$\chi$ direction described above with 
a circle in $\psi$ direction. The intervals of continuity have the 
structure of strips which run in
the $\psi$ direction around the domain. 
The natural domain of the higher dimensional scattering functions is
the product of the preimage of the line $C$ described above with the
circle of $\psi$ values. It is a two dimensional torus if we recall that
the initial and final point of the line $C$ should be identified.

Let us break the rotational symmetry and remember the
robustness mentioned at the end of the previous section.
For $A$ small the important intervals of continuity have still the same
qualitative structure, they are only deformed continuously. Only
the non hyperbolic dust around KAM islands is changed immediately
for $A \ne 0$. In this sense the qualitative structure of the
fractal of singularities is still close to a product of a one
dimensional fractal with a circle.

With increasing value of $A$ intervals of increasing 
size and lower level in the fractal hierarchy are changed 
qualitatively in so far as they are disrupted into
fragments which no longer have the structure of stripes winding
around the domain in $\psi$ direction. Only for $A \approx 0.5$ the
last large intervals of continuity are destroyed. Thereby the
last remnants of the product structure of the fractal are lost and
it is transformed into a truly higher dimensional fractal. In the 
present paper we shall not investigate this new structure, 
but restrict our consideration to smaller values of $A$.

So far we have seen that all topological 
information on the chaotic tangle is contained
in asymptotic scattering data, 
and that a measurement of scattering functions and
their analysis is one way for the asymptotic observer to obtain this 
information. The scattering functions present a kind of shadow image of the
chaotic invariant set cast into the outgoing asymptotic region.
However the measurement of scattering functions is difficult or 
impossible  in many
real scattering experiments and therefore we now have to go one step further 
to cross sections which are the quantity measured in most of the traditional
scattering experiments.

\section{The cross section}

The measurement of scattering functions needs the control over the
canonically conjugate variables $p$, $L$ and $\chi$, $\psi$. 
In classical mechanics
this could be done in principle. However, it is not done in the usual
scattering experiment. More important, in quantum mechanics this preparation
is forbidden even in principle, and if we want to develop concepts and
ideas which have some chance to be transferred to quantum dynamics,
 we can not use quantities needing the simultaneous 
preparation of canonically
conjugate variables. What is done in most scattering
experiment is the following: One half of the phase space variables (normally
the actions) is prepared as sharp as possible and the other conjugate half 
(the conjugate angles) is completely unspecified.
In our case this means: For fixed total energy $E$ we prepare $p_{in}$ 
and $L_{in}$ to  specific values and do not have any control 
over the reduced angles $\chi_{in}$ and $\psi_{in}$. Their values have 
a distribution with constant density over the whole domain, $\mathbb{T}^2$.
The detector measures the final values $p_{out}$ and $L_{out}$ for each 
outgoing particle and we monitor the relative probability $\sigma(p_{out}, \Delta L)$ 
to find a given combination of values of $p_{out}$ and $\Delta L$,
 normalised by the incoming flux.
This relative probability is called the doubly differential cross section. 
For more general information on cross sections 
see some text book on scattering theory as e.g. \cite{tay} or \cite{new}, 
for an application consult \cite{schelin}. 

In the previous section we have seen that the scattering function with values
$p_{out}(\chi_{in},\psi_{in})$, $\Delta L (\chi_{in}, \psi_{in})$ 
contains the fractal structure of the horseshoe.
If we can measure this function directly then we have
the necessary data to reconstruct 
the pattern of the horseshoe and our version
of the inverse problem is solved \cite{jls}. If we can only 
measure cross sections we must find out how the fractal pattern of the 
scattering function is transferred 
to some recognisable pattern in the cross 
section.

In the beam of incoming projectiles the weight on
the $\chi_{in},\psi_{in}$ torus is constant and the scattering function 
maps this initial weight on some interval 
in the range of this function having
the coordinates $p_{out}$ and $\Delta L$ and the image weight is
exactly the cross section $\sigma(p_{out},\Delta L)$. 
Therefore to get the value of
$\sigma$ for a particular combination of values of 
$p_{out}$ and $\Delta L$  
we first search for all preimages $\chi_k, \psi_k$ of 
the image point. Each preimage gives the contribution
\begin{equation}\label{weight}
g_k(p_{out},\Delta L) = 1/|Det \partial (p,\Delta L)/ \partial(\chi,\psi)|.  
\end{equation}
Then the value of the 
cross section is 
just the sum of these weights 
over all the contributing preimage points, i.e:
\begin{equation}\label{cross01}
\sigma(p_{out},\Delta L) = \sum_k g_k(p_{out},\Delta L)
\end{equation}

From equation \ref{weight} we see that the cross section 
has singularities at image points
where the Jacobian of this map is zero, i.e. 
points which are locally not invertible. They are caustics of the
projection of the graph of the function into the image space. 
The corresponding singularities in the cross section are 
called rainbow singularities, 
for their effect on light scattering off water drops. 
Now we shall focus on these singularities. 
Each interval of continuity of the scattering functions leads to one copy of
a typical rainbow singularity in the cross section. This gives a possibility 
to see the fractal structure 
of the chaotic invariant set in the cross section. 
For systems with two degrees of freedom this idea has been worked out in
detail in \cite{pue, schelin} and in this section we explain the higher dimensional
generalisation.

The determinant of the derivatives of the scattering function is zero
exactly for such values of $p_{out}$ and $\Delta L$ 
for which two trajectories
fall together and disappear. Accordingly these singularities are the lines
across which the number of contributing trajectories, i.e. the number of
preimages changes by 2. Now we will construct a simple analytical normal
form for these singularity lines and compare it with numerical results for 
the three model systems presented in section 2.

First we need an analytical model for the scattering function in one interval
of continuity which is as simple as possible but still produces the generic
form of singularity in the cross section. As explained in
section 4, the domain of the scattering function is the 2 dimensional torus   
with coordinates $\chi_{in}$ and $\psi_{in}$. 
For small deviation from symmetry a
typical interval of continuity is a strip running around the torus in $\psi$
direction. In $\chi$ direction the strips run in the symmetric case over a
limited range only, let us say from $\chi_0 - \delta$ up to $\chi_0 + \delta$,
where $\chi_0$ is the position of the middle of the strip. The value of
$p_{out}$ is maximal in the middle of the interval, let us say it has the
value $p_0$ in the middle, and it goes to zero on the boundaries of the
interval of continuity. Accordingly the simplest model function for $p_{out}$
in the symmetric case is
\begin{equation}\label{pout}
p_{out} = p_0 - (\chi_0 - \chi_{in})^2
\end{equation}
where the variables are scaled such that $\delta^2 = p_0$. For the asymmetric 
case we add a $\psi$ dependent term. This term must be $2 \pi$ periodic and for
simplicity we take the lowest Fourier contribution only, resulting in
\begin{equation}\label{poutcos}
p_{out} = p_0 - (\chi_0 - \chi_{in})^2 + b \cos(\psi_{in})
\end{equation}
for the general case with broken symmetry. Here for each given value of $\psi$
the variable $\chi$ is restricted to such values which result in a positive
value for $p_{out}$. This restriction defines the deformed interval of
continuity for the case of broken symmetry.

The function $\Delta L$ is identical zero in the symmetric case. For the
case of broken symmetry we will consider first the very simple model function
\begin{equation} \label{deltal}
\Delta L = a \sin(\psi_{in})
\end{equation}
without $\chi$ dependence and the version with $\chi$ dependence
\begin{equation}\label{deltalfactor}
\Delta L = a \sin(\psi_{in}) / (1 - c (\chi_{in} - \chi_0))
\end{equation}

The small perturbative parameters $b$ and $c$ 
will be considered as of the 
same order.

\begin{figure}
  \centering
  \subfigure[$b=0,c=0$]{\label{cata1}\includegraphics[width=0.45\textwidth]{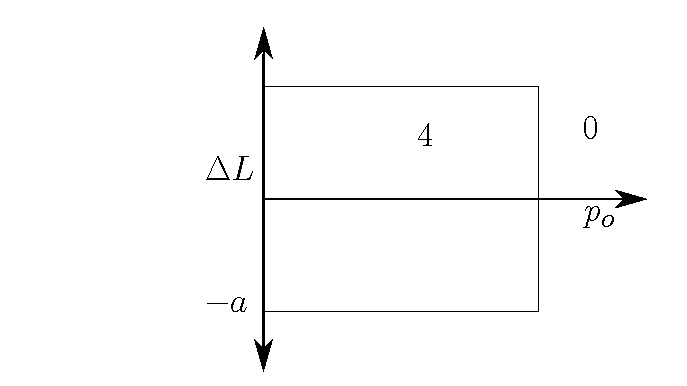}}   
  \subfigure[$b \neq 0,c=0$]{\label{cata2}\includegraphics[width=0.45\textwidth]{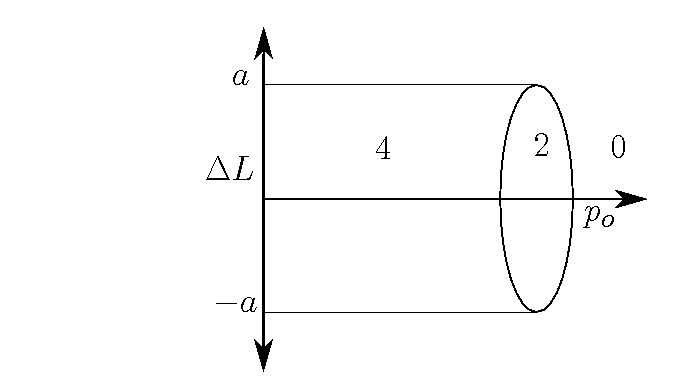}}
  \subfigure[$b=0, c\neq 0$]{\label{cata3}\includegraphics[width=0.45\textwidth]{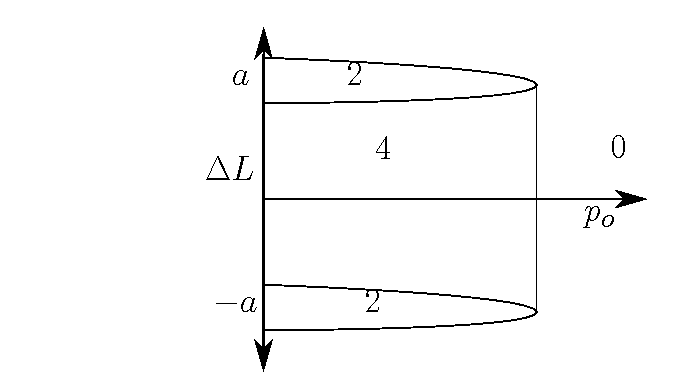}}
  \subfigure[$b\neq 0, c \neq 0$]{\label{cata4}\includegraphics[width=0.45\textwidth]{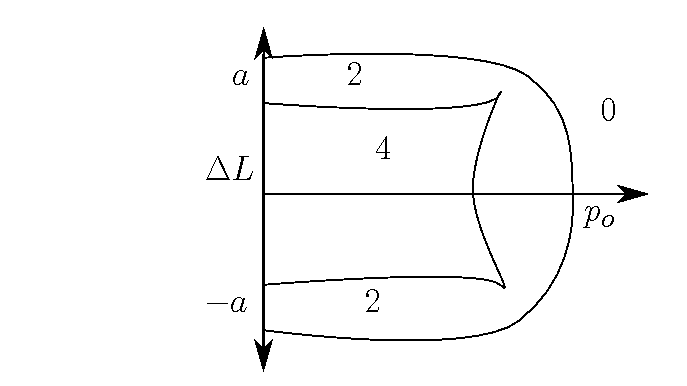}}
  \caption{The diagrams of the regions divided by a rainbow singularity. The numbers indicate
the quantity of preimages of the region for each point.}
  \label{diagzeros}
\end{figure}

Now we consider the preimages of these scattering model functions. First, as
simplest possibility combine Eqs. \ref{pout} and \ref{deltal}. In 
equation \ref{pout} we find 2 possible real
preimage values of $\chi$ as long as $p_{out} < p_0$ and 0 real preimage
values if $p_{out} > p_0$. Exactly at  $p_{out} = p_0$ two preimages collide
and turn from real to imaginary, as physical solutions they disappear.
Accordingly the line $p_{out} = p_0$ is a caustic line in this case.
In equation \ref{deltal} we find 2 real values of $\psi$ for $| \Delta L | < a$ and zero
preimages for $| \Delta L | > a$. Along the lines $ \Delta L  = a$ and
$ \Delta L  = -a$ the two solutions collide and turn from real to imaginary.
Therefore also these two lines are caustic lines for this simple case.
In total we find 4 preimages inside of the rectangle delimited by the lines
$p_{out}=0$, $p_{out} = p_0$, $ \Delta L  = a$ and $ \Delta L  = -a$ and
zero preimages outside of this rectangle (see figure \ref{diagzeros}).
Of course, the caustic structure of this very simple case is not structurally
stable, it changes qualitatively under small deformations of the scattering
functions. Along generic caustic lines the number of solutions changes by 2
and not by 4. Therefore we need to add the appropriate perturbations to the
scattering functions to turn the caustic structure into a structurally stable
one. We have to see next that the transitions from 
equation \ref{pout} to equation \ref{poutcos} and from
equation \ref{deltal} to equation \ref{deltalfactor} 
are the appropriate perturbations to turn the caustic lines
into generic and structurally stable ones.

Next, let us combine Eqs.\ref{poutcos} and \ref{deltal}.
In this case the function for $\Delta L$
still produces the degenerate caustic lines $ \Delta L  = a$ and
$ \Delta L  = -a$. If we invert equation \ref{poutcos}
 and we eliminate the $\psi$ dependence
by inserting from equation \ref{deltal} 
then we obtain for $\chi$ the equation
\begin{equation}
\chi_{in} = \chi_0 \pm \sqrt{ p_0 - p_{out} \pm b \sqrt{ 1 - \Delta L^2 / a^2}}  
\end{equation}
First we see the caustic lines $ \Delta L  = a$ and $ \Delta L  = -a$ which
are already known from the inversion of equation \ref{deltal}.
 In addition we also see that
solutions collide and turn from real to complex along the
ellipse given by the equation
\begin{equation}
a^2 (p_0 - p_{out})^2 + b^2 \Delta L^2 = a^2 b^2  
\end{equation}
In total for the combination of equations \ref{poutcos} and \ref{deltal}
 the behaviour of the preimages
is the following: Outside of the strip delimited by the lines
$ \Delta L  = a$ and $ \Delta L  = -a$ the number of real preimages is zero.
Let us now look at the interior of this strip. For values of $p_{out}$ 
outside  the ellipse and at the side of larger values we find zero real
preimages in $\chi$, inside the ellipse we find 2 preimages and for values
of $p_{out}$ between zero and the ellipse we find 4 preimages. The result is:
The perturbation introduced in equation \ref{poutcos}
 is able to modify the degenerate caustic
line $p_{out} = p_0$ into a structurally stable curve. However, the two
caustic curves $ \Delta L  = a$ and $ \Delta L  = -a$ stay in their
structurally unstable form.

Next let us combine Eqs.\ref{pout} and \ref{deltalfactor}. 
The equation\ref{pout} has the caustic line  
$p_{out} = p_0$ as before. Inserting equation\ref{pout} into 
equation\ref{deltalfactor} leads to the following
equation for $\psi$
\begin{equation}
\sin(\psi_{in}) = \Delta L (1 \pm c \sqrt{p_0 - p_{out}} ) / a, 
\end{equation}
which in turn leads to the caustic curves
\begin{equation}
\Delta L = \pm a (1 \pm c \sqrt{p_0 - p_{out}})^{-1}
\end{equation}
In the relevant region of $p$ values the curves given by this equation come 
close to parabolas centred at $\pm a$. Now the preimage
structure is the following. We find zero real preimages for $p_{out}>p_0$.
Therefore let us  concentrate on the strip of $p_{out}$ values between
zero and the line $p_{out} = p_0$. In the region of small values of 
$\Delta L$ between the two approximate parabolas we find 4 real preimages, 
inside of the two approximate parabolas
we find 2 real preimages and in the rest of the strip zero real preimages.
The perturbation introduced in equation\ref{deltalfactor}
 is able to turn the degenerate caustic
lines $ \Delta L  = a$ and $ \Delta L  = -a$ into structurally stable curves.

Last, let us check that the combination of Eqs.\ref{poutcos},
 \ref{deltalfactor}  produces structurally
stable caustic curves only. Eliminating the variable $\psi_{in}$ and making the
substitution $x = \chi_0 - \chi_{in}$ we get the following polynomial in $x$
\begin{equation}\label{Px}
P(x) = x^4 + A x^2 + B x + C
\end{equation}
where
\begin{equation}
A = 2 (p_{out} - p_0) + b^2 \Delta L^2 c^2 / a^2
\end{equation}
\begin{equation}
B = - 2 b^2 \Delta L^2 c / a^2
\end{equation}
\begin{equation}
C = (p_{out}-p_0)^2 + b^2 ( \Delta L^2 / a^2 -1)
\end{equation}
In the following we use the abbreviation $D =  \Delta L^2 / a^2 -1 $.
A polynomial has colliding solutions whenever its discriminant is zero, see
e.g. section II.4 and in particular lemma 4.7 in \cite{ken}. In the case of
the polynomial of  \ref{Px} the discriminant is given by
\begin{equation}
\operatorname{Dis} = -27 B^4 -4 A^3 B^2 + 16 A^4 C + 144 A B^2 C - 128 A^2 C^2 + 256 C^3
\end{equation}
We will evaluate the resulting complicated expression in $p$ and $x$
perturbatively in the small quantities $b$ and $c$. The lowest order
contributions come in order 4 and are
\begin{equation}
\operatorname{Dis}_4 = 256 b^4 (p_{out} - p_0)^2 D^2
\end{equation}
This discriminant is zero along the lines $p_{out} = p_0$, $\Delta L = a$
and $\Delta L = - a$. All these 3 lines come with multiplicity 2.
It is the same caustic structure as we obtained with the combination of 
the Eqs.\ref{pout} and \ref{deltal}.

\begin{figure}
  \centering
  \subfigure[]{\label{past1}\includegraphics[width=0.45\textwidth]{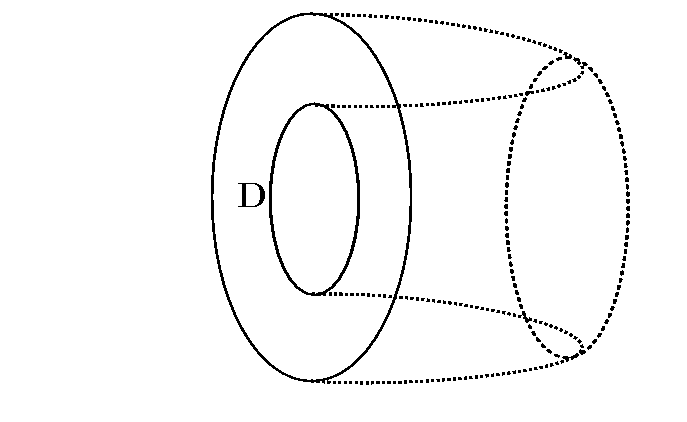}}   
  \subfigure[]{\label{past2}\includegraphics[width=0.45\textwidth]{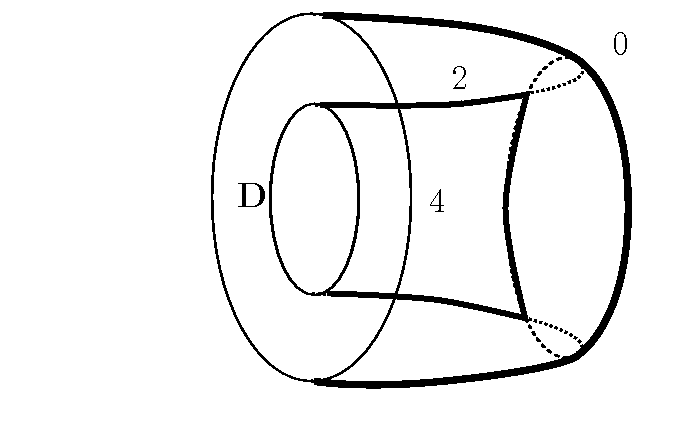}} 
  \caption{On the left side, labelled \ref{past1}, 
we represent an annular region which spans a quadratic maximum. 
On the right side, \ref{past2}, 
we emphasise the singular set of the projection into a two dimensional manifold. 
The domain of the Scattering function is labelled $D$. 
The numbers represent the number of inverse preimages for the projection. 
Compare with figures \ref{diagzeros} and \ref{singlert}.}
  \label{pastel01}
\end{figure}

\begin{figure}
  \centering
\subfigure[Rainbow Singularity for the channel system.]{\label{singlerh}\includegraphics[width=0.33\textwidth]{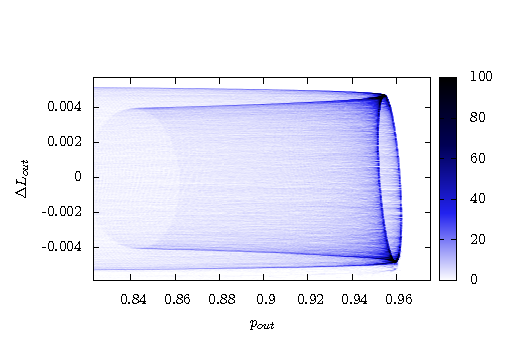}} 
  \subfigure[Rainbow Singularity for the map.]{\label{singlerm}\includegraphics[width=0.42\textwidth]{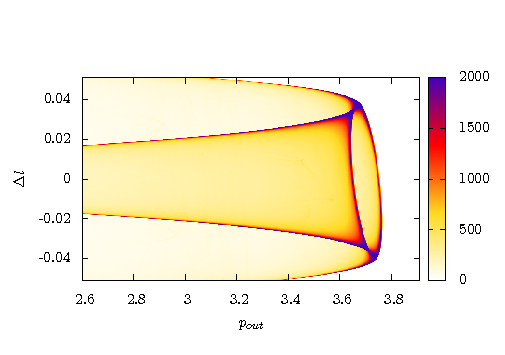}} 
\subfigure[Rainbow Singularity for the billiard.]{\label{singlerb}\includegraphics[width=0.42\textwidth]{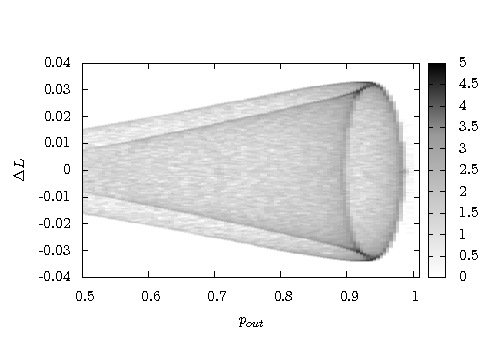}}     
  \subfigure[As a comparison, the zeros of the polynomial $Dis_6$, equation \ref{dis6}. The parameters are $a=1, b=c=0.1$]{\label{singlert}\includegraphics[width=0.42\textwidth]{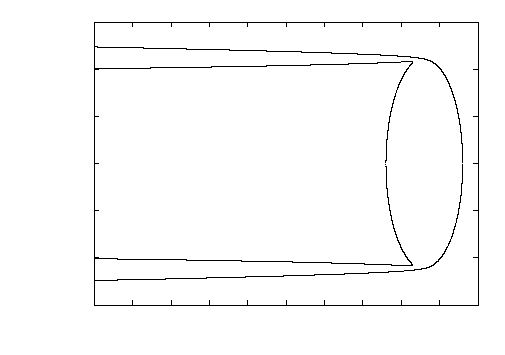}} 
  \caption{Different isolated rainbow singularities corresponding in each case 
 to a single domain of continuity of the scattering functions, 
for the different systems presented and the solution 
for the zeros of the $Dis_6$ polynomial.}
  \label{singles}
\end{figure}

Next we keep all contributions up to order 6 combined in $b$ and $c$ and
leave out the irrelevant constant factor $256 b^4$. The result is
\begin{equation}\label{dis6}
\operatorname{Dis}_6 = b^2 D^3 + (p_{out}-p_0)^2 D^2 + 2 (p_{out}-p_0)^3 ( \Delta L^2 
/ a^2 +1)\Delta L^2 c^2 / a^2 
\end{equation}
The real zeroes of these last expressions in the $(p_{out}, \Delta L)$ plane 
are sketched in
figure \ref{diagzeros}. We have also produced a theoretical plot for the curve
of the zeros of the expression \ref{dis6}  with  
 the parameter $a$  set on the value 1 whereas
the small parameters $b$ and $c$ take the value 0.1, see 
the figure \ref{singlert}. In these plots the various
regions of the plane are marked by the number of real preimages of the
original equations \ref{poutcos} and \ref{deltalfactor}. 
Here all structural instabilities of the caustic 
curves are removed. Note that this figure coincides with the projection
singularities of the ring shaped mountain shown in figure \ref{pastel01}.
The projection is highlighted on the left side. 
The parts a, b, c of figure \ref{singles} show for comparison the
rainbow structure in the cross section coming from a single interval of
continuity of the scattering function for the 3 examples introduced in section
2. Part a belongs to the channel with obstacle, part b
 to the map model and part c to the bottle billiard.

We see: For all the 3 examples the rainbow singularity in the cross section
has the same qualitative structure as the curves described by \ref{dis6}.
This motivates us to propose \ref{dis6}  as normal form of the rainbow singularity
in the double differential cross section for the class of scattering systems
considered in this article. 

The shape of the normal form induces a ring shaped mountain.
Since a ring shaped mountain can be considered half a torus, our projection 
is half of the well known projection singularity pattern of the torus, see
fig. 6.13 of \cite{ozorio}.

For moderate breaking of the symmetry each interval of continuity, which is
still some strip running around the torus in $\psi$ direction, produces in
the cross section one copy of the typical rainbow singularity. Of course,
for each individual interval it takes 
different values of the parameters. The total set
of singularities seen in the cross section coming from all intervals is
expected to consist of a superposition of a fractal of shifted and continuously
deformed copies of the normal form. In principle there should be an infinity 
of them. Practically we can resolve a finite number up to some finite
level of hierarchy of the underlying fractal. Let us turn next to some
numerical examples for the cross section.

\begin{figure}
\begin{center}
\includegraphics[width=0.75\textwidth]{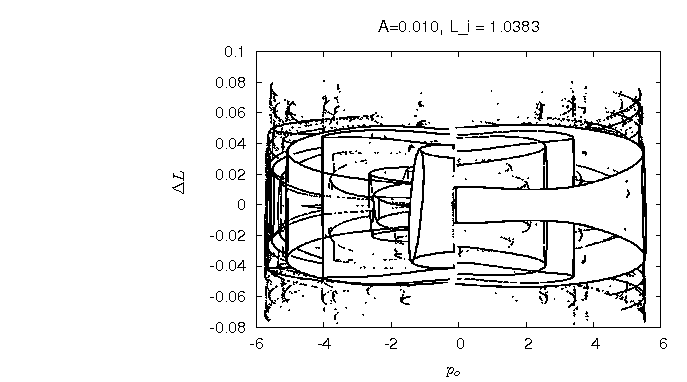}
\caption{The rainbow singularities for the map system generated by equation \ref{genmap}. The parameters are $A=0.010, L_{in}=1.0383$. We show the structure of the projection 
of the half torus repeated as far as resolution goes.}\label{seccexp01}  
\end{center}
\end{figure}

\begin{figure}
\begin{center}
\includegraphics[width=0.75\textwidth]{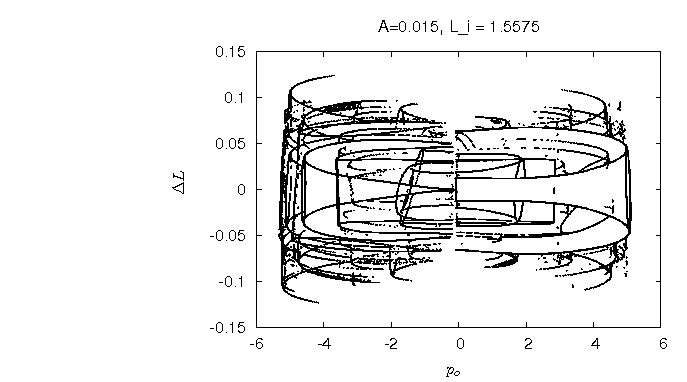}
\caption{The rainbow singularities for the system generated by the function $G(q,\theta,\tilde{p},\tilde{L})$, (equation  \ref{genmap}). The parameters are $A=0.015, L_{in}=1.5575$.}\label{seccexp02}   
\end{center}
\end{figure}

To obtain the figures 
\ref{singles}, \ref{seccexp01} and \ref{seccexp02} 
we have done a coarse graining of the domain of the cross section which is similar to
what any real detector in an experiment does. The domain is divided into
many small rectangles and the counts in any one of these rectangles are registered.
Each rectangle can be considered one detector channel. High count rates are indicated in the figure by
a darker colour. Note that there are high count rates also in some detector channels which
do not contain contributions from the true singularity itself, in particular the detector channels near 
the cusp of the inner singularity line show this behaviour. They are the ones roughly connecting
the inner cusp with the outer singularity line and marked by broken lines in figure \ref{past2}. The high count
rates lie along curves which come close to one ellipse and two parabolas, compare with figure \ref{pastel01}.

\begin{figure}
  \begin{center}
  \includegraphics[width=0.85\textwidth]{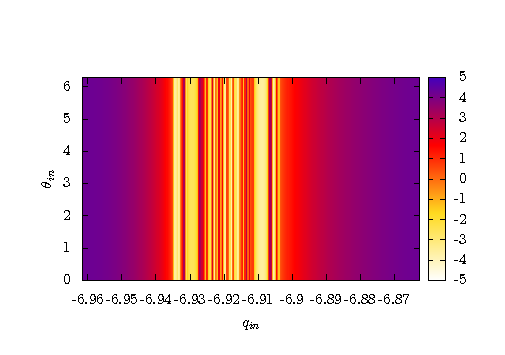} 
  \caption{The scattering function for the map generated by the function in expression \ref{genmap} with $A=0.000$ and $l_{in}=1.0383$ on the torus of initial conditions. As $L$ is conserved, we only show the $p_{out}$ component.}
  \label{tororoto01}
  \end{center}
\end{figure}

\begin{figure}
  \centering
  \subfigure[The $p_{out}$ component.]{\label{stripesp2}\includegraphics[width=0.85\textwidth]{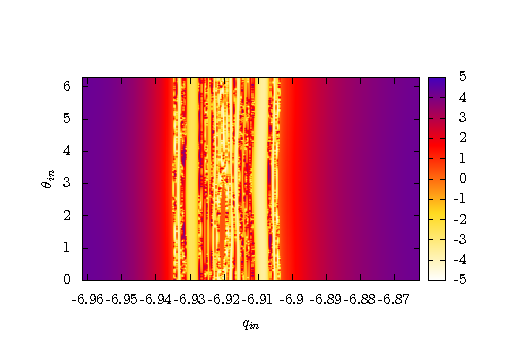}}   
  \subfigure[The $\Delta L$ component.]{\label{stripesl2}\includegraphics[width=0.85\textwidth]{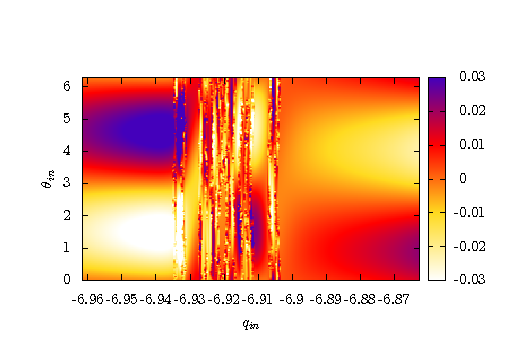}} 
  \caption{The scattering function for the map generated by the function in expression \ref{genmap} with $A=0.010$ and $L_{in}=1.0383$ on the torus of initial conditions. This gives the rainbow singularities depicted on figure \ref{seccexp01}.}
  \label{tororoto02}
\end{figure}

\begin{figure}
  \centering
  \subfigure[The $p_{out}$ component.]{\label{stripesp3}\includegraphics[width=0.85\textwidth]{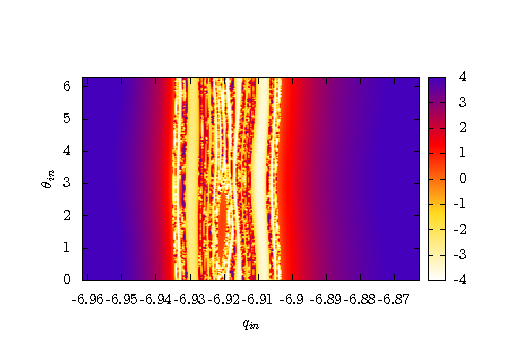}}   
  \subfigure[The $\Delta L$ component.]{\label{stripesl3}\includegraphics[width=0.85\textwidth]{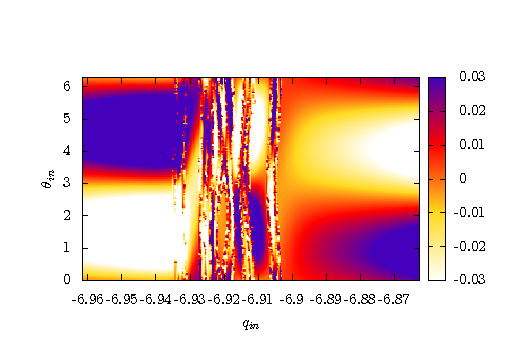}} 
  \caption{The scattering function for the map generated by the function in expression \ref{genmap} with $A=0.020$ and $L_{in}=1.0383$ on the torus of initial conditions.}
  \label{tororoto03}
\end{figure}

After studying the projections of these points 
we recognise the shape of the half toroidal mountains, see figures 
\ref{seccexp01}, \ref{seccexp02}. These figures have been obtained
by simulating a great quantity of trajectories with initial
conditions uniformly distributed 
on the torus. Then we have selected
points on the histogram of final conditions
 which show a much higher count than
their neighbours, as seen on the figures \ref{seccexp01} and \ref{seccexp02}.
 If we compare them with the black 
outline on the figure 
\ref{pastel01}, we can see the projection of the same basic
repeated over. For a
symmetric case we can see the domains of 
continuity in smoothly changing colours in the figure \ref{tororoto01}, which
partitions the torus into  a fractal family of stripes.  Each has
 typically an extremal set of measure zero and shows up
 as a
rainbow singularity
 in the cross section. 
As the symmetry gets broken, the embedding remains stable, 
although the discontinuities 
bend and may form rings (see figs. \ref{tororoto01} and \ref{tororoto02}), or
even multiply punctured domains.  
Then we observe distinct shapes besides the circular rings which should also 
represent a complete partition of the domain. It should be noted that the 
smoother parts of the function have still a ring shaped domain of continuity,
and that they are the main contributors to the cross section. 
Therefore, the experimentally detectable signature of the 
partitions of the domain
is still a family of ring shaped sets.

\section{Final remarks}

By asymptotic measurements we observe which rainbow contributions are present
or absent and can conclude which corresponding intervals of continuity in the
scattering function contribute and finally draw conclusions about the structure
of the chaotic invariant set in the Poincar\'e map of the system. In particular
we can change a parameter of the system, for example the symmetry breaking 
parameter A, and follow important changes of the chaotic set, at least in the
lower levels of the hierarchy of the fractal structure. For further
analysis of the resulting data two promising possibilities exist.

First, we can try to construct a symbolic dynamics for the system. For systems
with two degrees of freedom a description of the development scenario of the
chaotic set in terms of a development parameter related to an approximate symbolic
dynamics has been presented for binary and for ternary symmetric horseshoes in
\cite{jls} and \cite{rj1} respectively. This description was based on a rather
simple approximation for the symbolic dynamics. In the meantime for
chaotic sets of two dimensional maps, more sophisticated approximations for the
symbolic dynamics have been developed \cite{mide,mitch}. It would be worth to
generalise all these developments for the four dimensional maps encountered in
the present article.

Second, we can measure the weight of the contributions to the structures in
cross sections and scattering functions coming from the various levels of
hierarchy of the underlying fractal. Thereby we extract scaling factors of this
fractal. The distribution of these scaling factors can be analysed by thermodynamical
methods to extract the statistical measures of the chaos of the systems. For the
thermodynamical methods see \cite{tel,bec}. For the application of these methods
to chaotic scattering with two degrees of freedom and its cross section 
see \cite{tej}.

By knowing which rainbow singularities are present and which ones are
missing compared to the complete case and by knowing the scaling factors
between contributions from the various levels of the hierarchy we have
knowledge about the topology of the horseshoe and about the measures
of chaos in two dimensional cases.

All our considerations have been restricted so far to the case
where only one degree of freedom is open and all other degrees of
freedom are closed. This situation implies the following property
of the system: For any value of the total energy the closed degrees
of freedom can swallow all energy such that for the open degree
of freedom only energy zero remains. Therefore for any positive
value of the total energy the motion of the open degree of freedom 
can come to a stop at infinity ( when the potential is attractive 
without outer barrier ) or at the outermost potential barrier ( in
cases where it exists, where the potential is repulsive for large
distances ). In this sense the existence of at least one closed degree of 
freedom implies that for any positive value of the total energy
the system sits exactly on a channel threshold. As a consequence
we have the dividing surfaces of dimension 3 ( in the domain of the
Poincar\'e map ) formed by such
trajectories and the intersection of the stable and unstable
dividing surfaces forms the chaotic set described before.
The elementary intersection set between two hyper-surfaces of
codimension 1 is a set of codimension 2. If the dividing surfaces
form fractal folds, then we have a fractal collection of elementary
intersection patterns and the complete intersection set has a
codimension less than 2. In our particular case this property
guarantees that the chaotic invariant set in the Poincar\'e map has
a dimension larger than 2.

The situation is different if no closed degree of freedom exists. 
Imagine a system with only open degrees of
freedom and an attractive potential. Then for positive values
of the total energy it is impossible to have trajectories which
go out arbitrarily far and where at the same time the velocity goes
to zero. However, such trajectories are exactly the ones which form
the homoclinic tangle in the case investigated in the present paper
 and cause the fractal structure in the
scattering functions and in the cross section. This consideration
shows that the structures described in this article will not 
necessarily exist in systems without closed degrees of freedom.

Of course, topological chaos can exist in systems with only
open degrees of freedom. One system of this type with 3 degrees
of freedom, investigated in the past \cite{chen,korsch1}, is scattering
of a point particle off four hard spheres situated at the four
corners of a tetrahedron. When the radius of the spheres is small
compared to their distance then all periodic orbits are completely unstable
and we have hyperbolicity. In the 4 dimensional Poincar\'e map this
implies the existence of hyperbolic periodic points but excludes
the existence of NHIMs. The invariant manifolds of the hyperbolic
points are 2 dimensional, their elementary intersection structures
are points. Even when the manifolds are folded to form some fractal
structure, the complete intersection structure between stable and
unstable manifolds is a fractal built up of points. Then in general
the dimension of this intersection structure is small and does not
cause much observable structure in scattering functions.
According to \cite{chen} the intersection structure in the 4 dimensional
domain of the Poincar\'e map should have at least dimension 2 in
order to cause easily 
observable effects in generic scattering functions. 

The choice of our class of systems has also favourable consequences for
the cross sections. Naturally we look at the doubly differential
inelastic cross section as function of two action type variables.
In section 5 we studied the cross section as function of angular momentum
transfer and of outgoing momentum of the open degree of freedom.
Because of conservation of total energy and because of the
monotonous dependence of the energy of the oscillating degree of
freedom on its action we could equally well express the same cross
section as function of angular momentum transfer and the final action
of the oscillating degree of freedom. The use of the outgoing momentum 
has the advantage that its sign indicates immediately whether a 
particular trajectory describes transmission or reflection.
Therefore we stay with the variable $p_{out}$ in the cross section.
However, because of the above mentioned considerations we treat
$p_{out}$ as if it would be an action variable. Because of energy conservation
these two action like variables of the cross section are restricted 
to a finite interval of possible values. On the boundaries of
intervals of continuity of the scattering function $p_{out}$ goes
to zero. In the interior of intervals of continuity the scattering function
is smooth. Then this function necessarily has generic extrema in the
interior and has lines along which the determinant of derivatives
is zero. This guarantees the existence of the rainbow structures
described above.

Again the situation can be different for scattering systems with
open degrees of freedom only. For simplicity think of the scattering
of a points particle off a localised potential in a 3 dimensional
position space. For the moment consider the rotationally symmetric case
where the azimuth angle is irrelevant. Then the possibility exists that
in each interval of continuity of the scattering function the deflection 
angle goes monotonically from minus infinity to plus infinity without
any generic extremal points. As has been pointed out in \cite{moura} this
situation also can happen in cases of chaotic systems. Then the fractal
chaotic set in the phase space does not leave fractal traces in the
cross section. 

What happens for even more degrees of freedom, let say N? 
We can make some comments
for the class of systems where we have one open degree of freedom coupled
strongly to one closed degree of freedom and any number of further
degrees of freedom which are only coupled weakly to the first two degrees
of freedom. Let us assume again that there is a parameter $A$ which gives
the strength of this coupling and that we have again a limiting case
$A=0$ where these additional degrees of freedom are decoupled from the first
two ones and for simplicity also decoupled among themselves and where accordingly
the system can be reduced to one with two degrees of freedom. In this case
we find a (2N-4) dimensional NHIM in the (2N-2) dimensional domain of
the Poincar\'e map. It is stable in one direction, unstable in one direction
and neutrally stable in the remaining (2N-4) directions. The development degree
of the horseshoe of the reduced system depends on the amount of energy in the
reduced system. Since it depends on the value of the conserved actions of
the remaining degrees of freedom we can use one of the remaining neutral
directions as development parameter of the horseshoe and in total obtain
again a pile of two dimensional horseshoes similar to the 3 dimensional
tangle described before for the three degrees of freedom case. In particular
we can apply the same arguments of robustness as before. The difference
is that now we have to form the Cartesian product of this three dimensional 
structure with
(2N-5) neutral directions, which are partially action directions and
partially angle directions. The domain of the scattering function is now
a (N-1) dimensional torus of initial relative phase shifts between the N degrees 
of freedom and its range is a (N-1) dimensional interval of final actions.
Take our previous quantities $L$, $\theta$, $\chi$ and $\Delta L$
to be (N-2) component quantities.
Under small breaking of the symmetric case $A=0$ the robustness argument
indicates again that the coarse structure of the 
scattering function should be stable,
the lower the level of hierarchy in the fractal structure, the more stable
the situation should be.
The (N-1)-fold differential cross section is the relative probability to obtain
some combination of final actions for a constant density of the initial phase
shifts. The rainbow singularities are again the projection singularities of
the graph of the scattering function under projection on the range. The
elementary rainbow structure coming from a single interval of continuity of
the scattering function is now a (N-2) dimensional surface in the (N-1)
dimensional domain of the cross section. 
We plan a more detailed discussion of the general case in a future publication.

So far everything has been explained for classical dynamics. 
Therefore the question remains how our results are reflected in
quantum systems, like the ones presented in
\cite{mitchellbec}. We have analysed the cross section as function
of the outgoing action for fixed total energy. This was 
appropriate since classically the actions of the closed degrees
of freedom are continuous variables. In this point quantum mechanics
is essentially different. In our models asymptotically the open
degree of freedom is decoupled from the two closed degrees of
freedom, see for example equation \ref{hamgen2}.
 Accordingly in the asymptotic region
the transverse motion, i.e. the state of the two closed degrees
of freedom, must be in one of the discrete quantum states of this
bound subsystem. For example in equation \ref{hamgen2}
 the transverse
motion is a two dimensional harmonic oscillator with
its usual quantization of the action according to $I = (n+1/2)
\hbar \omega$. For other particular models of the closed transverse
degrees of freedom similar restrictions apply. On the other hand
this discreteness of the asymptotic transverse states provide a
natural channel decomposition of the S-matrix and the scattering 
amplitude where we start from a particular asymptotic initial 
transverse state and study the transitions to the various other
energetically allowed final asymptotic transverse states. 
The open longitudinal degree of freedom does not have similar 
restrictions, its asymptotic energy can be any positive value.
Thus to scan the action of the closed degrees of freedom is 
impossible in quantum dynamics. The appropriate procedure is to
select particular initial and final states of the closed degrees of
freedom and to vary the total energy
continuously, i.e. study the cross sections for the various
channel to channel transitions as function of the total energy.
In these quantities we expect to see structures, in particular
sequences of scattering resonances, related to the classical chaotic set.
So far we did not yet study such implications and they are not the
subject of the present publication.

The essential point for the analysis of the quantum mechanical cross section 
is the wave dynamical analog to classical rainbows.
One essential difference between classical dynamics and quantum dynamics is
how the contributions from the various initial conditions are summed.
In classical dynamics the cross section of 
equation \ref{cross01} itself is a sum over the
contributing initial conditions. In a semi-classical approximation to 
quantum dynamics or when using Feynman's path integral version of the 
quantum propagator, we first 
form a scattering amplitude as a sum over contributing trajectories and then
form the cross section or scattering probability as absolute square of the
amplitude. Therefore the resulting cross section is a double sum over the 
trajectories and contains first a sum over diagonal terms which has the 
structure of the classical cross section plus the non diagonal terms which are
interferences between the various contributing trajectories. If in a rainbow 
two trajectories fall together then in a semi classical treatment of the
scattering amplitude we have to 
apply some uniformization which replaces the contribution of the two
trajectories by an Airy function. This gives the wave dynamical 
analog of the classical square root singularity.

Accordingly two things might be done in a wave dynamical treatment. First,
the interference pattern in the wave cross section can be analysed. For
some attempts in this direction see \cite{tej} and \cite{jq1}, where the 
semi-classical scattering from three soft potential mountains has been used 
as example
of demonstration. Second, a decomposition of the scattering amplitude into
Airy type contributions might be tried, in order to recover at least the
first few hierarchical levels of the fractal pattern of rainbows. To our
knowledge this has not yet been done so far, but might be worth to try.

\section*{Acknowledgement}

This work has been supported by CONACyT under grant number 79988 and by DGAPA
under grant number IN-111607,
 and a doctoral scholarship grant also from CONACyT  for K. Zapfe. 
We thank Kevin Mitchell for useful comments on the manuscript.

\section*{References}

\bibliography{articulobib}

\end{document}